\documentclass[english,PRE,twocolumn]{revtex4-1}
\usepackage[T1]{fontenc}
\usepackage[latin9]{inputenc}
\usepackage{amstext}
\usepackage{amssymb}
\usepackage{graphicx}
\usepackage{wasysym}
\usepackage{esint}



\usepackage{babel}
\begin{document}

\title{Giant fluctuations in logistic growth}

\author{Bahram Houchmandzadeh}

\affiliation{CNRS, LIPHY, F-38000 Grenoble, France~\\
Univ. Grenoble Alpes, LIPHY, F-38000 Grenoble, France}
\begin{abstract}
We analyze the fluctuation of the number of individuals when two competing
species, beginning with a few initial individuals, are submitted to
a logistic growth. We show that when the total number of individuals
reaches the carrying capacity, the number of each species is subject
to giant fluctuations (variance $\sim$ mean$^{2}$) if the two species
have similar growth rate. We show that the deterministic logistic
equation can be used only when the growth rates are significantly
different, otherwise such growth has to be investigated by stochastic
processes tools. These results generalize to a wide class of growth
law. 
\end{abstract}
\maketitle

\section{Introduction.}

In many chemical or biological systems, fluctuations can be large
and drastically modify the results expected from a mean field approximation\cite{tsimring2014noisein}.
A famous early example was investigated by Delbrück \cite{delbruck1940statistical}
for the unbounded autocatalytic chemical reaction $A\rightarrow2A$
where he showed that the number $n(t)$ of $A$ molecules at time
$t$ displays giant fluctuations: the variance $V(t)$ is of the order
of the \emph{square} of the mean $V(t)=\left\langle n(t)\right\rangle ^{2}/n_{0}$,
where $n_{0}$ is the initial number of $A$ molecules. It can be
shown that spatial diffusion is not fast enough to dilute these local
fluctuations and this phenomenon can lead to spatial clustering for
example of organisms in ecological systems\cite{houchmandzadeh2008neutral,houchmandzadeh2009theoryof}
or of neutrons in nuclear reactors\cite{dumonteil2014particle}. 

The unbounded autocatalytic reaction captures the initial growth period,
but may seem unrealistic for systems where resources are limited\cite{das2012giantnumber}.
More realistic scenarios are captured by a logistic growth where the
reaction constant tends toward zero as the number of replicating agents
increases. If only one species is subject to such a growth, fluctuations
will become negligible when the number of replicating agents reaches
the carrying capacity of the system. On the other hand, as we show
below, if different species are competing for the same resources,
the number of each species can display large fluctuations similar
to the above example. This situation is relevant for example when
independent cellular pathways compete for the same resources\cite{genot2012computing},
when a cell is infected initially by a few bacteria or viruses carrying
different mutations or when different mutants of cancerous cells compete
with each other in the organism\cite{sprattdecelerating,atuegwu2013parameterizing}.
Another important example is chemical/biological reactions in small
compartments such as droplets\cite{song2006reactions,baccouche2017massively}
which can be used for example for high throughput directed evolution\cite{agresti2010ultrahighthroughput}. 

Consider the simple competition of two species of autoreplicators
$A$ and $B$ subject to a logistic growth where their deterministic
evolution equation is given by 
\begin{eqnarray}
\frac{dn}{dt} & = & an(N_{s}-n-m)\label{eq:det1}\\
\frac{dm}{dt} & = & bm(N_{s}-n-m)\label{eq:det2}
\end{eqnarray}
where $n$, $m$ are the (continuous) number of each species, $a$,
$b$, their respective growth rate at small concentration and $N_{s}$
the carrying capacity of the system. The solution of the above equations
is given by 
\begin{equation}
\frac{n}{n_{0}}=\left(\frac{m}{m_{0}}\right)^{r}\label{eq:sol1}
\end{equation}
where $n_{0}$ and $m_{0}$ are the initial number of each species
and $r=a/b$ is the relative growth rate of $A$ in respect to $B$
species. The final number of each species is found by solving $n_{\infty}+m_{\infty}=N_{s}$
in combination with relation (\ref{eq:sol1}). In particular, for
the neutral case $r=1$, the final number of each species is explicitly
given by $n_{\infty}=pN_{s}$ and $m_{\infty}=(1-p)N_{s}$ where $p=n_{0}/(n_{0}+m_{0})$
is the initial proportion of $A$ species. 

Equations (\ref{eq:det1}-\ref{eq:det2}) are mean field approximations
of the discrete stochastic process given by the following rates:
\begin{figure}
\begin{centering}
\includegraphics[width=0.8\columnwidth]{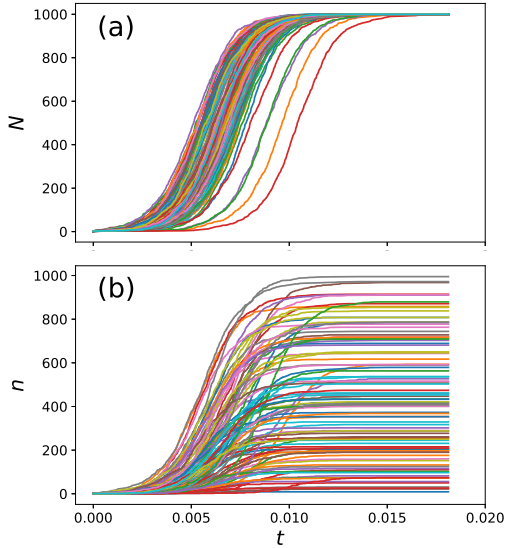}
\par\end{centering}
\caption{Neutral logistic growth of two competing species. 100 individual based
numerical simulation of equations (\ref{eq:sto1},\ref{eq:sto2})
with $a=b=1$ and $N_{s}=1000$ are displayed. The initial number
of each species is $n_{0}=m_{0}=1$. (a) Total number of individuals
$N$ ; (b) number of individuals of species $A$.  \label{fig:Neutral-logistic-growth}}
\end{figure}
\begin{eqnarray}
W(n,m\rightarrow n+1,m) & = & an(N_{s}-N)\label{eq:sto1}\\
W(n,m\rightarrow n,m+1) & = & bm(N_{s}-N)\label{eq:sto2}
\end{eqnarray}
where $N=n+m$ is the total number of individuals at time $t$. Figure
(\ref{fig:Neutral-logistic-growth}) displays the stochastic behavior
of the logistic growth (\ref{eq:sto1},\ref{eq:sto2}) for $r=1$.
We observe that as expected\cite{das2012giantnumber}, fluctuations
in the total number of individuals $N=n+m$ disappear as $N$ reaches
the carrying capacity $N_{s}$ (Fig. \ref{fig:Neutral-logistic-growth}a).
However, the number of individuals of each species is extremely variable
(Fig. \ref{fig:Neutral-logistic-growth}b). In fact, as we will show
below, the probability of finding $n$ individuals of type $A$ when
the system reaches saturation ($N=N_{s}$) is \emph{uniform} in this
case $P(n,N=N_{s})=1/(N_{s}-1)$. For such giant fluctuations, the
deterministic solution $n_{\infty}=pN$ is devoid of information and
we have as much chance of finding one $A$ individual as finding $pN$
individuals!

In this article, we investigate analytically and numerically the stochastic
equations (\ref{eq:sto1},\ref{eq:sto2}) in general and discuss the
origin of such large fluctuations when $r\approx1$. The following
section is devoted to the transformation of equations (\ref{eq:sto1},\ref{eq:sto2})
; section \ref{sec:neutral} investigates the problem for the neutral
case $r=1$ ; section \ref{sec:nonneutral} generalizes the solution
to $r\ne1$. The last section is devoted to discussion and concluding
remarks. Details of some computations are given in the appendices. 

\section{Mapping to a simple problem.}

Equations (\ref{eq:sto1},\ref{eq:sto2}) represent a 2+1 dimensional
system where because of the non-linearities, moment closure is lost
and no closed form solution can be obtained. However, if we change
the \emph{independent} variable from time $t$ to the total number
of individuals, the problem is mapped to a much simpler, one dimensional
one : instead of computing the probability $P(n,t)$ of finding $n$
individuals of type $A$ at time $t$, we compute the probability
$P(n,N)$ of finding $n$ individuals of type $A$ when the total
number of individuals is $N$. For long times, $N$ reaches the carrying
capacity $N_{s}$ and therefore, $P(n,t=\infty)$ and $P(n,N=N_{s})$
contain the same information. A similar transformation was recently~used
to compute the Luria-Delbrück distribution of the number of mutants
for a general growth curve\cite{houchmandzadeh2015general}. 

The Master equation governing $P(n,N)$ is simple. \emph{Once} a replication
event happens ($N\rightarrow N+1$), the probability that it was an
$A$ replicating $(n\rightarrow n+1)$ is 
\begin{eqnarray*}
\alpha_{N}^{n} & = & \frac{W(n,m\rightarrow n+1,m)}{W(n,m\rightarrow n+1,m)+W(n,m\rightarrow n,m+1)}\\
 & = & \frac{rn}{N+(r-1)n}
\end{eqnarray*}
The probability that it was a $B$ replicating ($n$ remains constant)
is 
\[
\beta_{N}^{n}=1-\alpha_{N}^{n}=\frac{N-n}{N+(r-1)n}
\]
The master equation for $P(n,N)$ is therefore 
\begin{figure}
\begin{centering}
\includegraphics[width=0.5\columnwidth]{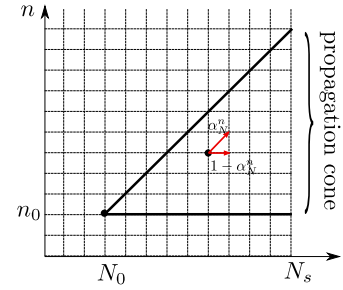}
\par\end{centering}
\caption{Mapping of the logistic growth into a flow problem in the $(N,n)$
plane.\label{fig:Mapping}}
\end{figure}
\begin{equation}
P(n,N+1)=\alpha_{N}^{n-1}P(n-1,N)+\left(1-\alpha_{N}^{n}\right)P(n,N)\label{eq:Master}
\end{equation}
at the initial time, the system contains $N_{0}$ individuals, $n_{0}$
of which are of type $A$ ; the initial condition for the Master equation
(\ref{eq:Master}) is 
\[
P(n,N_{0})=\delta_{n_{0}}^{n}
\]
where $\delta$ designates the Kronecker delta. The Master equation
(\ref{eq:Master}) is the mapping of the logistic growth into a flow
problem in the $(N,n)$ plane, where each node distributes its content
$P(n,N)$ to the adjacent ones $(N+1,n+1)$ and $(N+1,n)$ with proportion
$\alpha_{N}^{n}$ and $\beta_{N}^{n}$ (figure \ref{fig:Mapping}). 

Because of the form of the flow, the number of $A$ individuals $n$
is bounded by $n_{0}$ and $N-N_{0}+n_{0}$ (figure \ref{fig:Mapping}).
More over, on the two boundaries, the Master equation (\ref{eq:Master})
reduces to a one term recurrence relation. For example, on the lower
boundary, 
\begin{equation}
P(n_{0},N+1)=\left(1-\alpha_{N}^{n_{0}}\right)P(n_{0},N)\label{eq:lowerb}
\end{equation}
The probability is found to be 
\begin{equation}
P(n_{0},N)=\frac{(N_{0}-n_{0})_{N-N_{0}}}{(N_{0}+sn_{0})_{N-N_{0}}}\label{eq:lowerbsol}
\end{equation}
where $s=r-1$ is the excess relative fitness of species $A$. $(x)_{p}$
designates the Pochhammer symbol (raising factorial) :
\begin{equation}
(x)_{p}=x(x+1)...(x+p-1)\label{eq:pochhammer}
\end{equation}
Similarly, on the higher boundary, 
\begin{equation}
P(N-N_{0}+n_{0},N)=\frac{(n_{0})_{N-N_{0}}}{\left(\frac{N_{0}+sn_{0}}{r}\right)_{N-N_{0}}}\label{eq:higherbsol}
\end{equation}
Relation (\ref{eq:higherbsol}) can also be deduced from (\ref{eq:lowerbsol})
by exchanging the role of $A$ and $B$ individuals. 

The mean of various quantities can be computed theoretically from
the Master equation (\ref{eq:Master}). Let $f(.)$ be an arbitrary
function and define 
\[
\left\langle f(n)(N)\right\rangle =\sum_{n}f(n)P(n,N)
\]
then 
\begin{equation}
\left\langle f(n)(N+1)\right\rangle =\left\langle f(n)(N)\right\rangle +\left\langle \alpha_{N}^{n}\left(f(n+1)(N)-f(n)(N)\right)\right\rangle \label{eq:genmean}
\end{equation}
For example, for $f(n)=n$, we have 
\[
\left\langle n(N+1)\right\rangle -\left\langle n(N)\right\rangle =\left\langle \alpha_{N}^{n}\right\rangle 
\]
the mean field, continuous approximation of the above expression leads
to 
\begin{equation}
\frac{d\left\langle n\right\rangle }{dN}=\alpha_{N}^{\left\langle n\right\rangle }\label{eq:meanfieldmean}
\end{equation}
which is the equation deduced from the deterministic evolution (relation
\ref{eq:det1}-\ref{eq:det2}). 

Finally, note that it is very simple to compute numerically the probabilities
obeying the Master equation (\ref{eq:Master}): The right-hand side
of the equation (\ref{eq:Master}) is the product of a bi-diagonal
$(N+1)\times N$ matrix by an $N-$ column vector. 

The next two sections are devoted to the computation of the means
and probabilities for the neutral and non-neutral case.

\section{Solution for the neutral case.\label{sec:neutral}}

In the neutral case $r=1$, $\alpha_{N}^{n}=n/N$ ; the linearity
of $\alpha$ in $n$ allows for moment closure and efficient computation
of moments and probabilities. In particular, using relation (\ref{eq:genmean}),
the mean $\left\langle n(N)\right\rangle $ and variance $\sigma^{2}(N)$
are found to obey the recurrence equation
\begin{eqnarray}
\left\langle n(N+1)\right\rangle  & = & \left(1+\frac{1}{N}\right)\left\langle n(N)\right\rangle \label{eq:mean}\\
\sigma^{2}(N+1) & = & \left(1+\frac{2}{N}\right)\sigma^{2}(N)+p(1-p)\label{eq:var}
\end{eqnarray}
where $p=n_{0}/N_{0}$ is the initial proportion of the $A$ type.
The two first moments are then found to be 
\begin{figure}
\begin{centering}
\includegraphics[width=0.8\columnwidth]{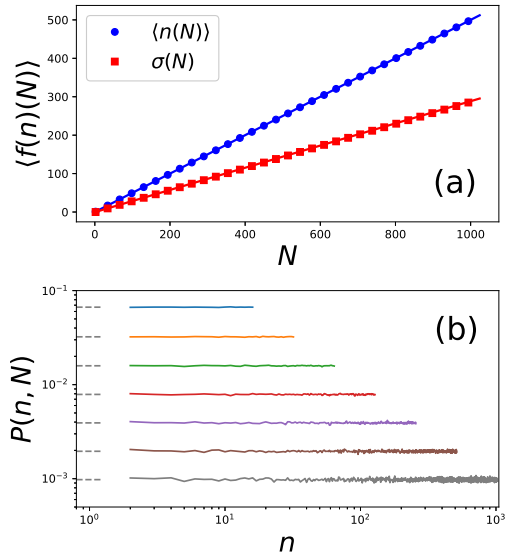}
\par\end{centering}
\caption{Numerical stochastic simulations of equations with rates (\ref{eq:sto1},\ref{eq:sto2})
and comparison to theoretical values for the neutral case $r=1$ and
initial values $N_{0}=2$ and $n_{0}=1$. (a) Evolution of the two
first moments $\left\langle n\right\rangle $ (circle) and $\sigma$
(squares) as a function of the number of individuals $N$. Symbols
: numerical stochastic simulations ; solid lines : theoretical values
given by relations (\ref{eq:meansolved},\ref{eq:varsolved}). The
moments were computed from $M=5000$ realizations. (b)Solid lines:
$P(n,N)$ as a function of the number of $A$ individuals $n$ for
various values of $N=2^{k}$, $k=4,8,...10$. The gray dashed lines
on the left designate the theoretical value $P(n,N)=1/(N-1)$ (relation
\ref{eq:n01N02}). The probabilities were computed from $M=10^{6}$
realizations\label{fig:MomentsProba}}
\end{figure}
\begin{eqnarray}
\left\langle n(N)\right\rangle  & = & pN\label{eq:meansolved}\\
\sigma^{2}(N) & = & \frac{p(1-p)}{N_{0}+1}N(N-N_{0})\label{eq:varsolved}
\end{eqnarray}
We observe that regardless of the population size $N$ of the system,
the fluctuations are of the same magnitude as the mean ($\sigma\sim\left\langle n\right\rangle $)
if the initial population size is small:
\begin{equation}
\text{cv}=\frac{\sigma(N)}{\left\langle n(N)\right\rangle }\approx\sqrt{\frac{1-p}{p(N_{0}+1)}}\label{eq:CV0}
\end{equation}
Figure (\ref{fig:MomentsProba}a) shows the perfect agreement between
stochastic numerical simulations (equations \ref{eq:sto1},\ref{eq:sto2})
and the above results on the moments.

Using expression (\ref{eq:genmean}), it can be shown (see appendix
\ref{subsec:Factorial-moments.}) that the raising factorial moments
obey a simple relation:
\begin{equation}
\left\langle (n)_{k}\right\rangle =\left\langle n(n+1)...(n+k-1)\right\rangle =\frac{(n_{0})_{k}}{(N_{0})_{k}}(N)_{k}\label{eq:factorialmoments-1}
\end{equation}

In the neutral case, we can go beyond moments computation and solve
the Master equation (\ref{eq:Master}) for $P(n,N)$. In general,
$P(n,N|n_{0},N_{0})$ is a polynomial of $n$ of degree $N_{0}-2$,
where $n_{0}$, $N_{0}$ are the initial conditions for the number
of $A$ individuals and all individuals. It is straightforward to
check that (see appendix \ref{subsec:Expression-of-the})
\begin{equation}
P(n,N|n_{0},N_{0})=A\frac{(n-n_{0}+1)_{n_{0}-1}(m-m_{0}+1)_{m_{0}-1}}{(N-N_{0}+1)_{N_{0}-1}}\label{eq:Pgeneralsol}
\end{equation}
where $m=N-n$, and by convention, $(x)_{0}=1$. The normalization
constant is found to be 
\[
A=\frac{(N_{0}-1)!}{(n_{0}-1)!(m_{0}-1)!}
\]
In particular, 
\begin{eqnarray}
P(n,N|1,2) & = & \frac{1}{N-1}\label{eq:n01N02}\\
P(n,N|2,3) & = & \frac{2(n-1)}{(N-1)(N-2)}\label{eq:n02N03}
\end{eqnarray}
The initial condition $n_{0}=1$, $N_{0}=2$ was used in numerical
simulations of figures \ref{fig:Neutral-logistic-growth},\ref{fig:MomentsProba}.
\begin{figure}
\begin{centering}
\includegraphics[width=0.8\columnwidth]{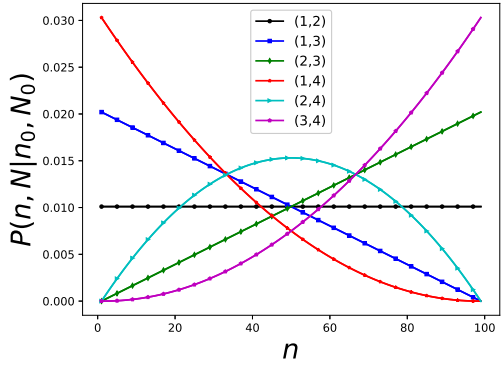}
\par\end{centering}
\caption{The probability $P(n,N|n_{0},N_{0})$ as a function of $n$ for $N=100$
and various initial conditions $(n_{0},N_{0})$ indicated in the legend.
Solid line : theoretical solution (\ref{eq:Pgeneralsol}) ; symbols
: numerical solutions of the master equation (\ref{eq:Master}). \label{fig:proba-IC} }

\end{figure}

The solution (\ref{eq:Pgeneralsol}) is in perfect agreement with
the numerical solution of the Master equation (\ref{eq:Master}) (figure
\ref{fig:proba-IC} ). 

\section{Solution for $r>1$.\label{sec:nonneutral}}

For the non-neutral case $r>1$, 
\[
\alpha_{N}^{n}=\frac{rn}{N+(r-1)n}
\]
is not anymore linear in $n$ and an exact solution for $P(n,N)$
becomes hard to obtain. However, as we are interested in the solution
for large $N$, we can treat $n$ and $N$ as \emph{continuous} variables
and approximate the Master equation (\ref{eq:Master}) by a partial
differential equation (PDE). The Master equation (\ref{eq:Master})
has indeed a simple structure and can be set into 
\begin{equation}
\partial_{N}P(n,N)+\partial_{n}\left[\alpha_{N}^{n}P(n,N)\right]=0\label{eq:MasterContinuous}
\end{equation}
Equation (\ref{eq:MasterContinuous}) is a first order PDE and can
be solved by the methods of characteristics\cite{polyanin2001handbook}.
Its general solution is found to be (see appendix \ref{sec:Solving-the-PDE})
\begin{equation}
P(n,N)=\frac{\partial}{\partial n}f\left(\frac{(N-n)^{r}}{n}\right)\label{eq:PDEsolgen}
\end{equation}
where $f(.)$ is an arbitrary function to be determined from the initial
condition. The implicit function $(N-n)^{r}/n=\text{Cte}$ is the
solution of the mean field equation (\ref{eq:meanfieldmean}) $dn/dN=\alpha_{N}^{n}$. 

\begin{figure}
\begin{centering}
\includegraphics[width=0.8\columnwidth]{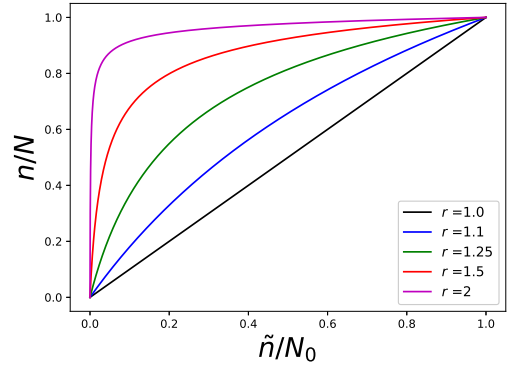}
\par\end{centering}
\caption{Function $n(\tilde{n})$ obtained by numerically solving the algebraic
equation (\ref{eq:implicitcondition}) for $N_{0}/N=10^{-3}$ and
various values of $r$. \label{fig:nnt} }

\end{figure}

Let us define $\tilde{n}$ such that (figure \ref{fig:nnt})
\begin{equation}
\frac{(N-n)^{r}}{n}=\frac{(N_{0}-\tilde{n})^{r}}{\tilde{n}}\label{eq:implicitcondition}
\end{equation}
Then for the initial condition $P(n,N_{0})=\phi_{0}(n)$, the complete
solution of equation (\ref{eq:MasterContinuous}) is given by (see
appendix \ref{sec:Solving-the-PDE})
\begin{eqnarray}
P(n,N) & = & \frac{\partial\tilde{n}}{\partial n}\phi_{0}(\tilde{n})\label{eq:PDEComplete}\\
 & = & \frac{\tilde{n}(N_{0}-\tilde{n})}{N_{0}+(r-1)\tilde{n}}\frac{N+(r-1)n}{n(N-n)}\phi_{0}(\tilde{n})\label{eq:PDECompleteb}
\end{eqnarray}
No special function is defined in the mathematical literature to deal
with equations of type $x^{r}+ux-u=0$ ; however, it is straightforward
to find the numerical solution of equation (\ref{eq:implicitcondition})
and use expression (\ref{eq:PDECompleteb}) to compute $P(n,N)$.

To make it more concrete, let us consider in some details the neutral
case $r=1$, and compare the exact known solution (\ref{eq:Pgeneralsol})
to the solution (\ref{eq:PDEComplete}) of the PDE approach. In this
case, relation (23) transforms into the explicit form $\tilde{n}=(N_{0}/N)n$.
The initial condition has to be chosen in order to match the known
solution (\ref{eq:Pgeneralsol}) ; once it has been fixed for $r=1$,
it will be used for all $r>1$. The initial condition corresponding
to the discrete case $n_{0}=1$, $N_{0}=2$ (relation \ref{eq:n01N02})
is 
\begin{figure}
\begin{centering}
\includegraphics[width=0.8\columnwidth]{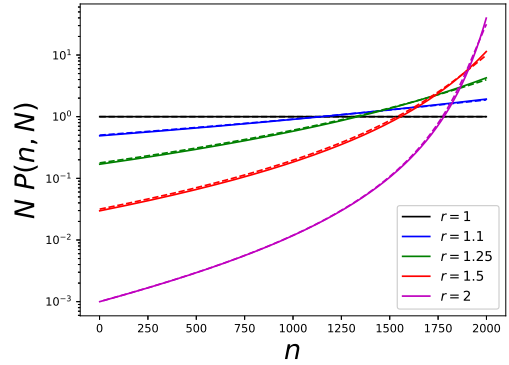}
\par\end{centering}
\caption{Solution (\ref{eq:PDEsolr}) of the continuous Master equation (\ref{eq:MasterContinuous})
(continuous lines) compared to numerical solutions of the discrete
Master equation (\ref{eq:Master}) (dashed lines) for $N=2000$, $N_{0}=2$,
$n_{0}=1$ and various values of $r$. The solution (\ref{eq:PDEsolr})
is obtained by numerically solving equation (\ref{eq:implicitcondition})
and then using relation (\ref{eq:PDECompleteb}). \label{fig:PnPDE}}
\end{figure}
\begin{equation}
\phi_{0}(n)=\Pi(n-1)\label{eq:phi0N2n1}
\end{equation}
where the gate function is defined as $\Pi(x)=1/2$ for $\left|x\right|<1$
and is zero outside this domain. Therefore, 
\begin{eqnarray}
P(n,N) & = & \frac{2}{N}\Pi\left(\frac{2}{N}n-1\right)\label{eq:PDEsolr1}\\
 & = & \frac{1}{N}\,\,\,\,\,\,\,\,\,n\in]0,N[
\end{eqnarray}
which approximates the exact solution (\ref{eq:n01N02}) to $O(1/N)$. 

The general solution for arbitrary $r$ corresponding to initial condition
$n_{0}=1$, $N_{0}=2$ is then simply
\begin{equation}
P(n,N)=\frac{1}{2}\frac{\partial\tilde{n}}{\partial n}\,\,\,\,\,\,n\in]0,N[\label{eq:PDEsolr}
\end{equation}
Figure \ref{fig:PnPDE} shows the excellent agreement between expression
(\ref{eq:PDEsolr}) and the numerical solution obtained from the exact
discrete Master equation (\ref{eq:Master}).

Various moments can be extracted from solution (\ref{eq:PDEComplete}):
\begin{equation}
\left\langle n^{k}(N)\right\rangle _{r}=\int_{0}^{N}n^{k}P(n,N)dn=\int_{0}^{N_{0}}n^{k}\phi_{0}(\tilde{n})d\tilde{n}\label{eq:momentsequation}
\end{equation}
where $n$ inside the integrand on the right-hand side of eq. (\ref{eq:momentsequation})
is a function of $\tilde{n}$ through relation (\ref{eq:implicitcondition}).
For the neutral case $r=1$, $n/N=\tilde{n}/N_{0}$ and therefore
\begin{equation}
\int_{0}^{N_{0}}\frac{\tilde{n}^{k}}{N_{0}^{k}}\phi_{0}(\tilde{n})d\tilde{n}=\frac{\left\langle n^{k}(N)\right\rangle _{1}}{N^{k}}=\frac{(n_{0})_{k}}{(N_{0})_{k}}+O(1/N)\label{eq:momentphi}
\end{equation}

We can obtain an explicit form of $n$ as a function of $\tilde{n}$
for various conditions. If $s=r-1\ll1$, we can obtain a perturbative
solution of equation (\ref{eq:implicitcondition}) in powers of $s$.
On the other hand, for high values of integer $r$ such as $r=2,3,4$,
we can exactly solve the algebraic equation (\ref{eq:implicitcondition}).
These two cases constitute the near neutral and highly non-neutral
situations and allows us to understand the general behavior of the
system. 

\subsection{Perturbative solution.\label{subsec:Perturbative-solution.}}

Let us first consider the case $s=r-1\ll1$. Setting $\kappa=\log(N/N_{0})$,
we have, to the second order in $s$ : 
\begin{equation}
x=\tilde{x}+\kappa\tilde{x}(1-\tilde{x})s-\kappa\tilde{x}(1-\tilde{x})\left((\kappa+1)\tilde{x}-\kappa/2\right))s^{2}\label{eq:xtperturbative}
\end{equation}
 where $\tilde{x}=\tilde{n}/N_{0}$, $x=n/N$. The symmetry of equation
(\ref{eq:implicitcondition}) implies that $\tilde{x}$ can be expressed
as a function of $x$ by simply replacing $\kappa$ by $-\kappa$
in expression (\ref{eq:xtperturbative}). Using expression (\ref{eq:momentsequation},\ref{eq:momentphi})
for the initial conditions $N_{0},n_{0}$, , to the first order perturbations,
the moments are found to be
\begin{figure}
\begin{centering}
\includegraphics[width=1\columnwidth]{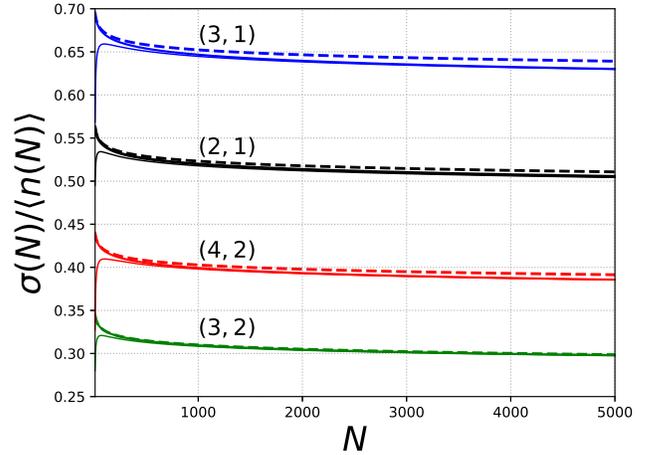}
\par\end{centering}
\caption{Coefficient of variation $\sigma(N)/\left\langle n(N)\right\rangle $
for $s=0.05$ ($r=1+s$) and various initial conditions $(N_{0},n_{0})$.
Thin solid lines : exact values obtained from direct numerical resolution
of the Master equation (\ref{eq:Master}) ; Dashed lines : first order
perturbations given by expression (\ref{eq:cvs}) ; thick solid lines
: second order perturbations. The initial condition $(N_{0},n_{0})$
of each curve is displayed above it.\label{fig:CVs}}
\end{figure}
\begin{eqnarray}
\left\langle n(N)\right\rangle _{r} & = & \left\langle n(N)\right\rangle _{1}\left\{ 1+\kappa s\frac{(N_{0}-n_{0})}{N_{0}+1}\right\} \label{eq:mu1s}\\
\sigma_{r}^{2}(N) & = & \sigma_{1}^{2}(N)\left\{ 1+2\kappa s\frac{N_{0}-2n_{0}}{N_{0}+2}\right\} \label{eq:vars}\\
\frac{\sigma_{r}(N)}{\left\langle n(N)\right\rangle _{r}} & = & \text{cv}_{1}\left\{ 1-\kappa s\frac{N_{0}(n_{0}+1)}{(N_{0}+1)(N_{0}+2)}\right\} \label{eq:cvs}
\end{eqnarray}
Where the subscript $1$ refers to the neutral expressions (\ref{eq:meansolved}-\ref{eq:CV0}).
Figure \ref{fig:CVs} shows the comparison of the above expressions
to exact values obtained from numerical solutions of the exact Master
equation (\ref{eq:Master}).

We observe that the correction of the above expressions compared to
neutral values (equation \ref{eq:meansolved}-\ref{eq:CV0}) are logarithmic
and of the order of $s\kappa=s\log(N/N_{0})$ : the fluctuations amplitude
$\sigma$ is still large and of the order of the mean $\left\langle n\right\rangle $.
The perturbative approach is valid for $\kappa s\ll1$ ; the solution
for higher values of $s$ can be slightly improved by using higher
order perturbations ( figure \ref{fig:CVs}) but the perturbative
approach reaches its limit for $\kappa s\lesssim1$. 

\subsection{High values of $r$.}

High values of $r$ can be understood by investigating integer values
such as 2,3,4 for which the equation (\ref{eq:implicitcondition})
can be exactly solved. For the case $r=2$ 
\begin{equation}
x=\frac{\gamma(1-\tilde{x})^{2}+2\tilde{x}-\sqrt{\gamma^{2}(1-\tilde{x})^{2}+4\gamma\tilde{x}}}{2\tilde{x}}\label{eq:xr2}
\end{equation}
where $\gamma=N_{0}/N\ll1$ and as before, $x=n/N$ and $\tilde{x}=\tilde{n}/N_{0}$.
We will investigate the simplest case corresponding to the initial
condition $N_{0}=2$, $n_{0}=1$ where $\phi_{0}(u)=\Pi(u-1)$ (relation
\ref{eq:phi0N2n1}). For this initial condition, the moments equation
(\ref{eq:momentsequation}) is greatly simplified :
\begin{equation}
\frac{\left\langle n^{k}\right\rangle }{N^{k}}=\frac{N_{0}}{2}\int_{0}^{1}x^{k}d\tilde{x}\label{eq:simplifiedmoments}
\end{equation}
Using expression (\ref{eq:xr2}), performing the integrations involved
by equation (\ref{eq:simplifiedmoments}) and keeping only the leading
orders of $\gamma$, we find that 
\begin{figure}
\begin{centering}
\includegraphics[width=0.9\columnwidth]{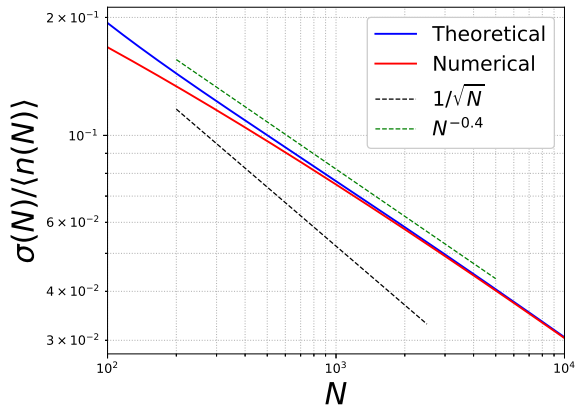}
\par\end{centering}
\caption{Coefficient of variation $\sigma(N)/\left\langle n\right\rangle _{N}$
for $r=2$ with initial condition $N_{0}=2$, $n_{0}=1$. The theoretical
value is obtained from expressions (\ref{eq:mu1r},\ref{eq:varr})
; the exact, numerical value is obtained by numerical resolution of
the Master equation (\ref{eq:Master}). As a guide for the eye, $N^{-0.5}$
and $N^{-0.4}$ are also displayed.\label{fig:CVr2}}

\end{figure}
\begin{eqnarray}
\frac{\left\langle n(N)\right\rangle }{N} & = & 1-\frac{4}{3}\sqrt{\gamma}+\frac{\gamma}{4}\left(1-2\log\gamma\right)+O(\gamma^{3/2})\label{eq:mu1r}\\
\frac{\sigma^{2}(N)}{N^{2}} & = & \gamma\left(-\log\gamma-\frac{77}{18}\right)+\nonumber \\
 &  & \gamma^{3/2}\left(-\frac{4}{3}\log\gamma+\frac{106}{15}\right)+O(\gamma^{2})\label{eq:sigma2r}
\end{eqnarray}
Expression (\ref{eq:sigma2r}) is valid for $N/N_{0}\apprge72$ which
is indeed the regime of interest (figure \ref{fig:CVr2}) . We see
that for $r=2$, the variance increases only as $N\log N$ and not
$N^{2}$ as in the neutral case. Therefore, for high values of $N$,
the coefficient of variation $\sigma/\left\langle n\right\rangle $
decreases as $(\log N/N)^{1/2}$ . In this regime, fluctuations become
negligible and the deterministic approach is valid. 

\section{Discussion and Conclusion.}

In this article, we have investigated the distribution of the number
of individuals $n$ and $m$ of two species $A$,$B$ during a logistic
growth. We have shown that the investigation is greatly simplified
if instead of time $t$, the independent variable is chosen to be
the total number of individuals $N=n+m$. This paper was focused on
the well known logistic growth, but the method and conclusions are
valid for any stochastic growth of the form
\begin{eqnarray}
W(n,m\rightarrow n+1,m) & = & anf(n,m)\label{eq:sto1-1}\\
W(n,m\rightarrow n,m+1) & = & bmf(n,m)\label{eq:sto2-1}
\end{eqnarray}
 where $f(n,m)$ is an arbitrary function not necessarily symmetric
in $m$ and $n$. 

The most interesting feature of the investigated system is the large
amplitude of fluctuations in the neutral case $r=a/b=1$, where both
species have similar growth rate. Suppose that we draw (and replace)
$N_{s}$ individuals at random from a pool of $N_{0}$ individuals
when $n_{0}$ are of the $A$ type. The distribution of the number
of $A$ type in the $N_{s}$ sample is a binomial one with parameter
$p=n_{0}/N_{0}$ ; the fluctuation amplitude of this experiment $\sigma/\left\langle n\right\rangle \sim1/\sqrt{N_{s}}$
is small if $N_{s}\gg1$. One could naively suppose that a logistic
growth when two types $A$ and $B$ individuals are competing and
the system expands from $N_{0}$ to $N_{s}$ individuals ($N_{s}\gg N_{0}$)
is similar to the above drawing experiment : each individual in the
final pool draws at random its ancestor from the initial pool. This
is however not the case and we have shown that contrary to the binomial
case, the fluctuation amplitude $\sigma/\left\langle n\right\rangle \sim1/\sqrt{N_{0}}$
is always large and independent of the final system size. 

Various experiments can be devised to test the relevance of the above
computations. For example, a phage such as $\lambda$ can be modified
into few different mutants, each expressing a different fluorescent
proteins (such as GFP, RFP, YFP,...) ; the mutants can then be used
to co-infect a bacterial culture. The distribution of the colors in
the culture after some time can be related to the probabilities we
have computed through a convolution by a Poisson-Binomial distribution
to account for variation in the initial number of co-infectors. A
similar experiment can be performed using PCR amplification of few
similar DNA strands\cite{schaerli2009continuousflow} of the same
length and characteristics and then analyze the number of strands
copy in each droplets.

The problem we have investigated can also be used to extend the Wright-Fisher
(WF) model of population genetics to variable size population (see
for example\cite{ewens1967theprobability,otto1997theprobability,wienand2017evolution}).
In the WF model with fixed population size $N_{0}$ and two mutant
types $A$ and $B$, each generation is formed by selecting randomly
$N_{0}$ individuals among the progeny of generation $i$ to form
generation $i+1$. If $x$ is the proportion of the $A$ type with
reproductive advantage $r=1+s$, then a diffusion (Kimura) equation
can be derived for the evolution of the population (\cite{kimura1955solution,ewens2004mathematical})
where the drift and diffusion coefficient are $a(x)=sx(1-x)$ and
$b(x)=x(1-x)/(2N)$. 

We can generalize the WF model by allowing, at each generation $i$,
the population to expand to size $N_{s}$ and then select $N_{0}$
individuals among them to form the new generation $i+1$. By using
the result of subsection \ref{subsec:Perturbative-solution.}, it
is straightforward to show that the diffusion equation governing this
system is the same as before except that the relative excess fitness
is now renormalized to $s'=s\log(N_{s}/N_{0})$. The fact that the
effective fitness increases in a growing population was already noted
by Ewens (\cite{ewens1967theprobability}), although the amplifying
factor in the problem investigated by him was proportional to the
harmonic mean $N_{s}$ and $N_{0}$ rather than their logarithmic
difference as here. 

In summary, we have shown that populations subjects to logistic-like
growth such as equation (\ref{eq:sto1-1},\ref{eq:sto2-1}) can be
modeled by deterministic equations only if there is significant difference
($r\apprge2$) between their growth rates. If they have similar growth
rate, the deterministic equation must be abandoned and a stochastic
treatment used instead.

\begin{acknowledgments}
I thank Luca Peliti, David Lacoste, Marcel Vallade, Alexandre Dawid and
Hidde De Jong for fruitful discussions and critical reading of this manuscript.
\end{acknowledgments}

\appendix

\section{Various neutral computations}

\subsection{Factorial moments.\label{subsec:Factorial-moments.}}

Consider the function $f(n)=(n)_{k}=n(n+1)...(n+k-1)$ ; then 
\[
f(n+1)-f(n)=(n+1)_{k-1}\left(n+k-n\right)=k(n+1)_{k-1}
\]
and therefore 
\[
n\left(f(n+1)-f(n)\right)=k(n)_{k}=kf(n)
\]
Therefore, using the general expression (\ref{eq:genmean}), we find
the one term recurrence relation 
\[
\left\langle f(n)(N+1)\right\rangle =\left(1+\frac{k}{N}\right)\left\langle f(n)(N)\right\rangle 
\]
which is trivially solved and leads to expression (\ref{eq:factorialmoments-1}). 

\subsection{Expression of the probability.\label{subsec:Expression-of-the}}

To shorten the notations, we use $m=N-n$ whenever needed. The Master
equation in the neutral case is 
\begin{equation}
P(n,N+1)=\frac{n-1}{N}P(n-1,N)+\frac{m}{N}P(n,N)\label{eq:masteralt}
\end{equation}
Consider 
\begin{equation}
P(n,N)=\frac{(n-n_{0}+1)_{n_{0}-1}(m-m_{0}+1)_{m_{0}-1}}{(N-N_{0}+1)_{N_{0}-1}}\label{eq:palt}
\end{equation}
Pochhammer manipulation is similar to factorial manipulation. In particular,
\begin{eqnarray*}
N(N-N_{0}+1)_{N_{0}-1} & = & (N-N_{0}+1)_{N_{0}}\\
(n-1)(n-n_{0})_{n_{0}-1} & = & (n-n_{0})_{n_{0}}\\
m(m-m_{0}+1)_{m_{0}-1} & = & (m-m_{0}+1)_{m_{0}}
\end{eqnarray*}
and therefore, the right hand side of relation (\ref{eq:masteralt})
is found to be 
\[
\frac{(n-n_{0}+1)_{n_{0}-1}(m-m_{0}+2)_{m_{0}-1}}{(N-N_{0}+1)_{N_{0}}}\left(n-n_{0}+m-m_{0}+1\right)
\]
As 
\[
n-n_{0}+m-m_{0}+1=N-N_{0}+1
\]
and 
\[
\frac{N-N_{0}+1}{(N-N_{0}+1)_{N_{0}}}=\frac{1}{(N+1-N_{0}+1)_{N_{0}-1}}
\]
expression (\ref{eq:palt}) is indeed a solution of the Master equation,
up to a multiplicative constant. The constant is found by stating
$P(n_{0},N_{0})=1$. As the master equation conserves the probability,
the constant is valid for all $N$. 

\section{Solving the PDE\label{sec:Solving-the-PDE}}

Consider a first order partial differential equation (PDE) of first
order for the function $P(x,t)$ of type
\begin{equation}
\partial_{t}P+\partial_{x}(\alpha P)=0\label{eq:pdesol}
\end{equation}
where $\alpha=\alpha(x,t)$ is a known function. Let us call $R(x,t)=\text{Cte}$
the solution of the characteristic equation 
\[
\frac{dx}{dt}=\alpha(x,t)
\]
Then by definition, 
\[
\partial_{t}R+\alpha\partial_{x}R=0
\]
Consider the function 
\begin{equation}
Q(x,t)=\frac{\partial}{\partial x}f\left(R(x,t)\right)\label{eq:mysol}
\end{equation}
where $f()$ is an arbitrary function. Then
\[
\partial_{t}Q+\partial_{x}\left(\alpha Q\right)=\partial_{x}\left\{ \left(\partial_{t}R+\alpha\partial_{x}R\right)f'(R)\right\} =0
\]
and therefore $Q(x,t)$ is a solution of equation (\ref{eq:pdesol}).
For example, for $\alpha=c$, the solution is the trivial propagation
$P(x,t)=f(x-ct)$. 

The function $f(.)$ has to be determined from the initial condition
$P(x,t_{0})=\phi_{0}(x)$. Consider two points $(t_{0},\tilde{x})$
and $(t,x)$ in the plane, related through $R(x,t)=R(\tilde{x},t_{0})$,
$i.e.$ they belong to the same characteristic curve. Obviously, we
can reverse this relation as $\tilde{x}=g\left(R(x,t),t_{0}\right)$
and therefore write the general solution (\ref{eq:mysol}) as $P(x,t)=\partial_{x}f\left(\tilde{x}\right)=(\partial\tilde{x}/\partial x)f'(\tilde{x})$.
On the other hand, at the initial time $t_{0}$, $x=\tilde{x}$, $\partial\tilde{x}/\partial x=1$
and therefore $f'()=\phi_{0}().$ The solution of the PDE (\ref{eq:mysol})
with the initial condition $\phi_{0}(x)$ is then 
\[
P(x,t)=\frac{\partial\tilde{x}}{\partial x}\phi_{0}(\tilde{x})
\]
$P(.,t)$ can be seen as a transformation, \emph{i.e. }scaling and
deformation of the initial condition $\phi_{0}(.)$. An initial Dirac
distribution however propagates without deformation along a characteristic
curve because $f(x)\delta(x)=\delta(x)$ : in this case, the PDE is
reduced to the deterministic equations $dx/dt=\alpha$. 

Let us precise the function $\phi_{0}(.)$ used in this article for
the PDE (\ref{eq:MasterContinuous}) . The true probability $P_{d}(n,N)$
is function of \emph{discrete} variables $n$ and $N$. In order to
estimate this probability, we have used the probability density $P_{c}(n,N)$
of \emph{continuous} variable $n,N$. $P_{c}$ must approximate $P_{d}$
for \emph{large} $N$. $\phi_{0}()$ has to be chosen to make this
approximation as precise as possible. However, we cannot use the discrete
initial condition $P(n,N_{0})=\delta_{n_{0}}^{n}$, because the continuous
PDE will be reduced to a deterministic equation. We make the assumption
that the choice of $\phi_{0}(.)$ is independent of $r$ and therefore
can be deduced from the known expression of neutral probability. For
$r=1$, $\tilde{n}=(N_{0}/N)n$, and therefore we have 
\[
\phi_{0}(\tilde{n})=\frac{N}{N_{0}}P_{1}(\frac{N}{N_{0}}\tilde{n},N)
\]
where $P_{1}(,)$ is the neutral probabilities but the arguments are
continuous.

\bibliographystyle{unsrt}

\end{document}